\documentclass[a4paper]{jpconf}
\usepackage{graphicx}
\begin{document}
\title{Thermodynamic curvature: pure fluids to black holes}

\author{George Ruppeiner}

\address{Division of Natural Sciences, New College of Florida, Sarasota, Florida 34243-2109}

\ead{ruppeiner@ncf.edu}

\begin{abstract}
Thermodynamics unavoidably contains fluctuation theory, expressible in terms of a unique thermodynamic information metric. This metric produces an invariant thermodynamic Riemannian curvature scalar $R$ which, in fluid and spin systems, measures interatomic interactions. Specifically, $|R|$ measures the size of organized fluctuating microscopic structures, and the sign of $R$ indicates whether the interactions are effectively attractive or repulsive. $R$ has also been calculated for black hole thermodynamics for which there is no consensus about any underlying microscopic structures. It is hoped that the physical interpretation of $R$ in fluid and spin systems might offer insight into black hole microstructures. I give a brief review of results for $R$ in black holes, including stability, the sign of $R$, $R=0$, diverging $|R|$, and various claims of ``inconsistencies'' in thermodynamic metric geometry.
\end{abstract}

\section{Introduction}
Thermodynamics \cite{Callen} has a reputation as being a science complete in its basic laws. Less well known is thermodynamic fluctuation theory which follows from thermodynamics \cite{Landau}. This fluctuation theory may be represented by a unique thermodynamic information Riemannian metric \cite{Ruppeiner1995}, resulting in an invariant Riemannian thermodynamic curvature scalar $R$. Physically, $R$ reveals information about interatomic interactions, and I illustrate here with recent calculations of $R$ in pure fluids with thermodynamic data from the NIST fluid database \cite{Ruppeiner2012}.

As was shown by Bekenstein \cite{Bekenstein}, Hawking \cite{Hawking}, and others, thermodynamics may be extended to black holes. Logically, thermodynamic fluctuation theory, including the physical interpretation of $R$ developed in fluid and spin systems, should extend likewise. This paper includes a tabulation of calculations of $R$ from general relativistic black hole solutions, for which no underlying microscopic models are known. I also discuss various claims of ``inconsistencies'' concerning divergences of black hole heat capacities, and cases with $R=0$.

\section{Pure fluids}
The basic structure in thermodynamic fluctuation theory in pure fluids is an infinite environment in some reference state ``0'', characterized by two independent intensive thermodynamic variables, and containing an open subsystem with constant volume $V$. This subsystem exchanges energy and particles with its environment, and its thermodynamic state fluctuates about an equilibrium characterized by maximum entropy. The probability of a fluctuation away from equilibrium is given by Einstein's famous formula \cite{Landau}: $\mbox{probability}\propto\mbox{exp}\left[-V(\Delta\ell)^2 /2 \right]$, where $(\Delta\ell)^2$ is the invariant, positive definite, thermodynamic information metric. Expressed in the pair of independent thermodynamic variables $x^1$ and $x^2$: $(\Delta \ell)^2 = g_{\mu\nu}\Delta x^{\mu}\Delta x^{\nu}$, where $\Delta x^\alpha\equiv (x^\alpha -x^\alpha_0)$ denotes the difference between the thermodynamic variables $x^{\alpha}$ of the subsystem and their equilibrium values $x^{\alpha}_0$. Denote by $S$, $X^1$, and $X^2$ the entropy, internal energy, and particle number, respectively, of the subsystem. $X^1$ and $X^2$ are each conserved quantities, and $S$ is additive between the subsystem and its environment. If $x^{\alpha}=X^{\alpha}$, then:

\begin{equation} g_{\alpha\beta} = -\frac{1}{k_B V}\frac{\partial^2 S} {\partial X^{\alpha}\partial X^{\beta}},\label{30}\end{equation}

\noindent where $k_B$ is Boltzmann's constant \cite{Ruppeiner1995}. $g_{\alpha\beta}$ transforms as a second-rank tensor.

The thermodynamic metric Eq. (\ref{30}) leads to an invariant Riemannian thermodynamic curvature scalar $R$. $R=0$ for the ideal gas \cite{Ruppeiner1979}, and $|R|$ diverges as the correlation volume $\xi^3$ near the critical point \cite{Ruppeiner1995,Ruppeiner1979,Johnston2003}. $R$ is thus a measure of effective interatomic interactions. Model calculations reveal that the sign of $R$ corresponds to whether interactions are effectively attractive ($R<0$) or repulsive ($R>0$); see the tabulation in \cite{Ruppeiner2010}. The connection of $|R|$ to fluctuating structure size has also been established directly by means of a covariant thermodynamic fluctuation theory \cite{Ruppeiner1995}.

Most recently \cite{Ruppeiner2012}, $R$ has been calculated in pure fluids using data based on experiment. Pure fluids have interatomic interaction potentials of the Lennard-Jones type, strongly repulsive at short range, attractive at long range, and with a minimum at a distance where the atoms in the condensed liquid and solid phases reside. A fluid typically has density smaller than that of a condensed state, and we thus expect an average attractive interatomic interaction, and negative $R$. Negative $R$ is indeed the norm in pure fluids. Figure 1 shows $R$ for hydrogen along the coexistence curve, with: 1) $R$ in both phases diverging to $-\infty$ at the critical point, 2) $|R|$ in both phases decreasing as the triple point is approached, with liquid phase values approaching the order of a molecular volume, and 3) a region of small positive $R$ in the condensed liquid phase near the triple point as the liquid organizes into a solid-like structure.

One may now pose a puzzle: which atomic arrangements produce the thermodynamic curvatures seen in Fig. 1? Figure 2 suggests a solution encompassing most fluid situations: Fig. 2a shows widely spaced atoms where attractive interactions form loose fluctuating clusters of volume $|R|\propto\xi^3$. Such structures are typical of the critical point regime. Fig. 2b shows a tight cluster of atoms in the vapor phase prevented from collapse by repulsive interactions. Such clusters are present in water, but not in hydrogen \cite{Ruppeiner2012}. Fig. 2c shows a haphazard arrangement of atoms in a condensed liquid state. Fig. 2d shows an organized solid-like state. The liquid phase in Fig. 1 shows a change in sign of $R$ corresponding to a transition Fig. 2c $\to$ Fig. 2d.

Results for $R$ in pure fluids are thus associated with simple microscopic patterns. Black hole physics does not have as well developed a theoretical microscopic structure. However, the calculation of $R$, and insights from fluid or spin models, might lend some guidance in this difficult domain.

\begin{figure}[h]
\begin{minipage}{18pc}
\includegraphics[width=18pc]{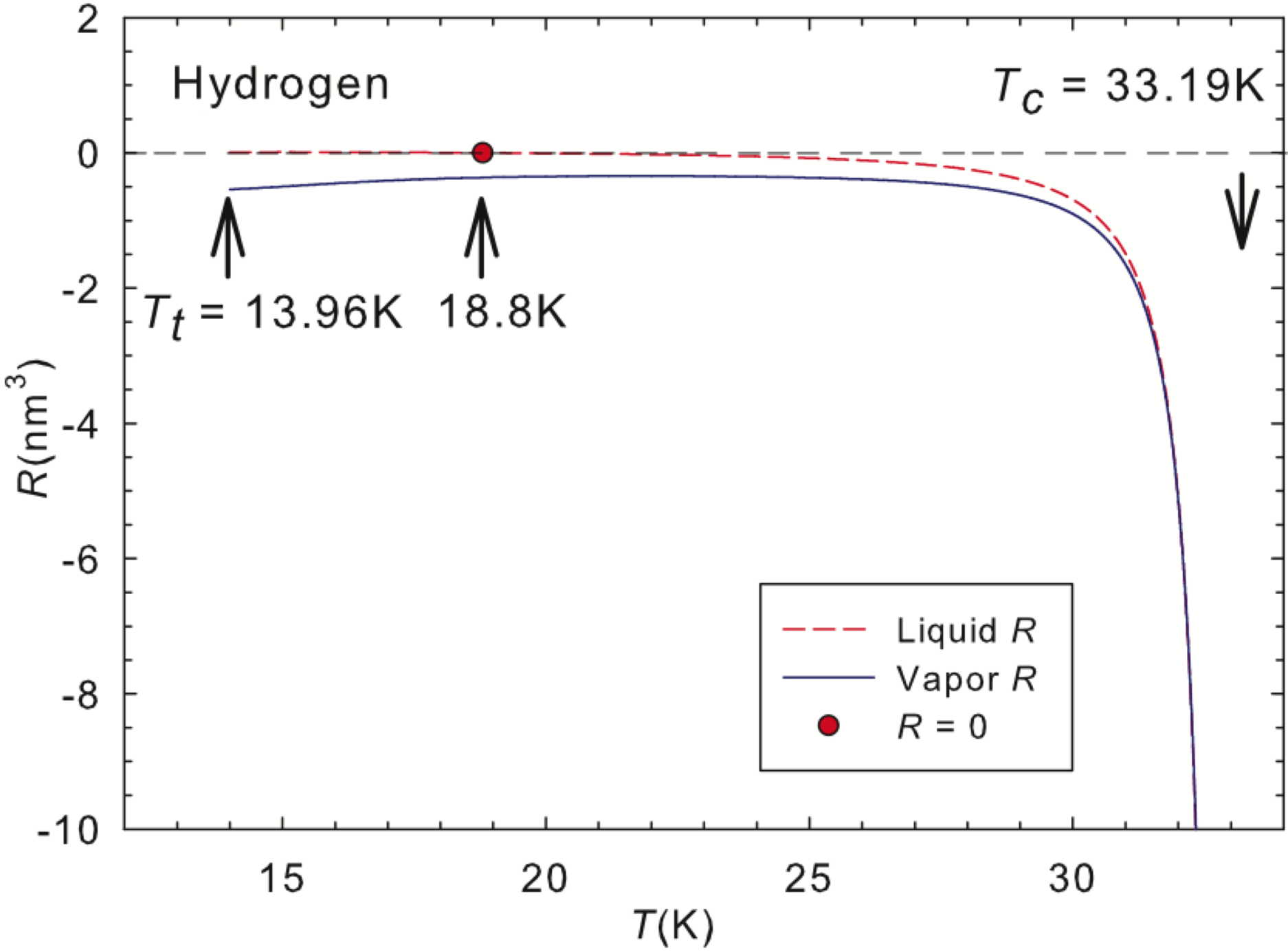}
\caption{\label{label}$R$ for hydrogen along the coexistence curve in the liquid and vapor phases from the triple point to the critical point, with temperatures $T=T_t$ and $T=T_c$.}
\end{minipage}
\hspace{0.4cm}
\begin{minipage}{18pc}
\includegraphics[width=18pc]{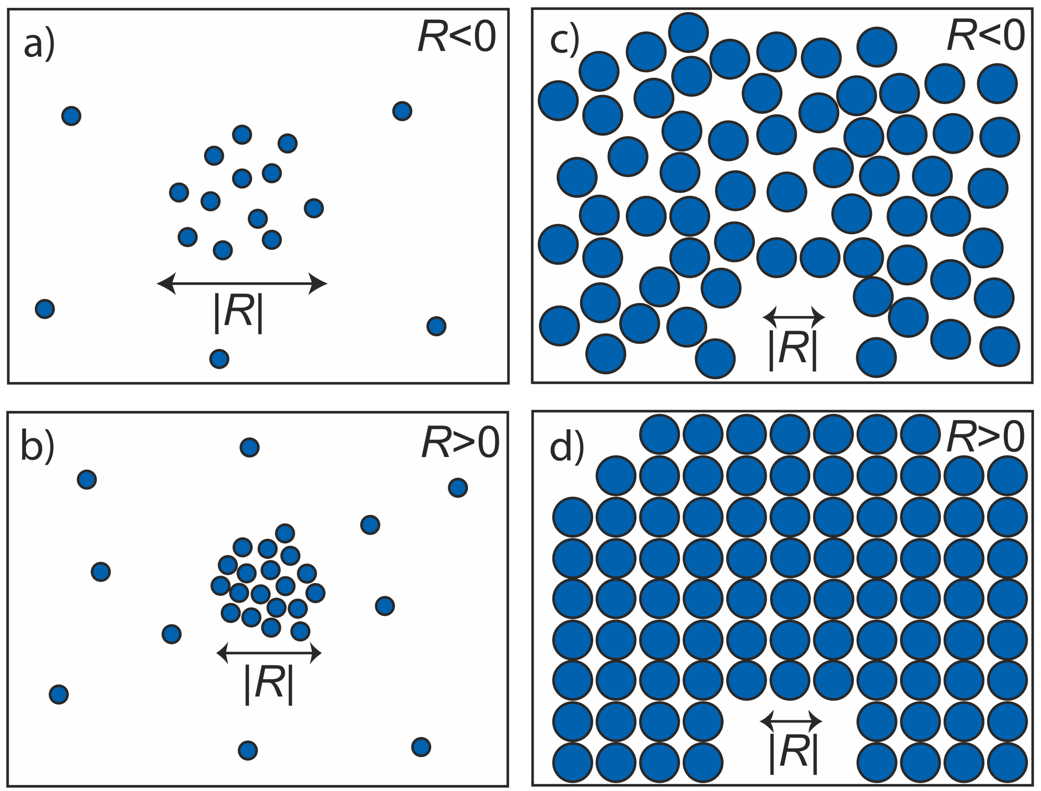}
\caption{\label{label} a) Atoms pulled together by attractive interactions; b) atom cluster held up by repulsive interactions; c) disorganized liquid phase; and d) organized solid phase.}
\end{minipage}\hspace{0pc}
\end{figure}

\section{Black holes}
Black hole solutions to general relativity have well-defined thermodynamic properties. The role of the conserved thermodynamic variables is played by mass, angular momentum, and charge $(M, J, Q)$, respectively. Logically, black hole thermodynamics leads to thermodynamic fluctuation theory and an information metric Eq. (\ref{30}) (set $V=1$, and use $(M, J, Q)$ for the $X^\alpha$'s \cite{Ruppeiner2008}). This information metric produces the black hole thermodynamic curvature $R$ \cite{Ferrara1997,Aman2003}. A tentative attempt to physically interpret $R$ for black holes has been made in terms of the spatial size of correlated fluctuations on the event horizon \cite{Ruppeiner2008}. Perhaps the puzzle solving tried above in pure fluids can eventually lead to insight about possible black hole microstructures.

Debated in black hole thermodynamics has been the possibility raised by Davies \cite{Davies1977} that the curve of diverging heat capacity $C_{J,\,Q}\equiv T(\partial S/\partial T)_{J,\,Q}$ in the Kerr-Newman black hole corresponds to a phase transition. Since diverging heat capacities are a feature of second-order phase transitions in fluid and spin systems, Davies' association would appear logical. However, in ordinary thermodynamics the connection between thermodynamic anomalies and phase transitions gets support from microscopic models. Without such models there can be no assurance of any phase transition. For black holes, if some heat capacity diverges, how could we be sure that we have not just made an inappropriate choice of thermodynamic variables, which reveals infinities with no really fundamental significance? What would we make of curves where one heat capacity diverges, but the other heat capacities stay regular? What if each heat capacity diverges along its own curve, as happens in the Kerr-Newman black hole \cite{Ruppeiner2008}? Which curve corresponds to a true phase transition? These questions identify difficulties associating diverging heat capacities with phase transitions. It seems simpler to look at diverging $|R|$ to indicate phase transitions, since $R$ has a unique status in identifying microscopic order, which in fluid and spin systems goes to the heart of the issue of phase transitions.

Black hole solutions with $R$ identically zero have also given rise to debate; see \cite{Medved2008} for a review. Examples of such solutions are the BTZ and the Reissner-Nordstr$\ddot{\mbox{o}}$m \cite{Aman2003} black holes. If $R$ measures in some sense the range of interactions, then one might expect $|R|$ to be large for black hole thermodynamics, reflecting the gravitational forces present in these objects. But such reasoning need not obtain. In a classical black hole, the gravitating particles have collapsed to a central singularity, shrinking the interactions between them to zero volume. The statistics underlying the thermodynamics might reside on the event horizon, where unknown constituents might interact with each other by forces perhaps not gravitational. In this scenario, gravity might merely be a nonstatistical force holding the assembly together, and a result $R = 0$, where the unknown constituents move independently of each other, might not be so unreasonable.

\begin{table}
\centering
\caption{\label{book} Properties of $R$ for black holes solutions from general relativity.}
\begin{tabular}{@{}l*{15}{c}}
\br
Name of solution								& Dimension		& Variables		& Stable	& Sign of $R$	& $R=0$	& $ \frac{|R|\to\infty}{T \ne 0}$	\\
\mr
BTZ \cite{Aman2003}							& $2 + 1$			& ($M$, $J$)		& yes	& $0$ 		& -		& no						\\
Kerr \cite{Aman2003}							& $3 + 1$			& ($M$, $J$)		& no		& $+$		& no		& no						\\
Reissner-Nordstr$\ddot{\mbox{o}}$m\cite{Aman2003}	& $3 + 1$			& ($M$, $Q$)		& no		& $0$		& -		& no						\\
Kerr-Newman \cite{Ruppeiner2008,Mirza2007}		& $3 + 1$			& ($M$, $J$, $Q$)	& no		& $+$		& no		& no						\\
K-AdS \cite{Sahay2010b}							& $3 + 1$			& ($M$, $J$)		& yes	& $-$		& no		& yes					\\
RN-AdS \cite{Aman2003}							& $3 + 1$			& ($M$, $Q$)		& yes	& $\pm$		& yes	& yes					\\
KN-AdS \cite{Sahay2010}							& $3 + 1$			& ($M$, $J$, $Q$)	& yes	& $\pm$		& yes	& yes					\\
Black Hole \cite{Arcioni2005}						& $4 + 1$			& ($M$, $J$)		& no		& $+$		& no		& no						\\
Small Black Ring \cite{Arcioni2005}					& $4 + 1$			& ($M$, $J$)		& no		& $\pm$		& yes	& yes					\\
Large Black Ring \cite{Arcioni2005}					& $4 + 1$			& ($M$, $J$)		& yes	& $-$		& no		& yes					\\
\br
\end{tabular}
\end{table}

Table 1 summarizes a number of black hole calculations of $R$. Tabulated are the spatial ($+$ time) dimensions, whether or not there are regimes of thermodynamic stability (positive definite metric matrix Eq. (\ref{30})), the sign of $R$ (or an indication ``0'' if $R$ is identically zero), whether or not there are places where the sign of $R$ changes through zero, which might indicate a Hawking-Page phase transition \cite{Sahay2010}, and whether or not there are divergences $|R|\to\infty$ with $T\ne 0$. Most solutions here with $R$ not identically zero diverge to $\pm\infty$ at the extremal curve $T\to 0$. I omit string theory black hole solutions for simplicity. Table 1 shows a progression from the elegant BTZ black hole, stable for all states, and with identically zero $R$, to the older solutions Kerr, Reissner-Nordstr$\ddot{\mbox{o}}$m, and Kerr-Newman, none of which rise to the level of stability. (Omitted are cases of the Kerr-Newman black hole in which one of the parameters is held fixed at value not zero, while the others fluctuate; such do show interesting instances of stability \cite{Ruppeiner2008}). A richer structure, including stability, $R$ zero crossing, and $R$ divergences, clearly results from either increasing the number of dimensions, adding an asymptotic AdS scenario, and/or more complex topology. Too intricate to tabulate here are results for Myers-Perry black holes \cite{Aman2010}.

\ack
I thank George Skestos for travel support, and Bhupendra Tiwari for productive discussions.

\end{document}